\documentclass{article}

\usepackage{latexsym}

\usepackage{amssymb}

\newcommand{\R}{\mathbb{R}}
\newcommand{\fc}{f_{\cap}}
\newcommand{\jc}{j_{\cap}}
\newcommand{\rhoc}{\rho_{\cap}}
\newcommand{\Ec}{E_{\cap}}
\newcommand{\Bc}{B_{\cap}}
\newcommand{\ec}{e_{\cap}}
\newcommand{\pc}{\mathfrak{p}_{\cap}}

\newcommand{\Uc}{U_{\cap}}
\newcommand{\phic}{\phi_{\cap}}
\newcommand{\psic}{\psi_{\cap}}
\newcommand{\fcin}{\fc^{\mathrm{in}}}
\newcommand{\Ecin}{\Ec^{\mathrm{in}}}
\newcommand{\Bcin}{\Bc^{\mathrm{in}}}
\newcommand{\phicin}{\phic^{\mathrm{in}}}
\newcommand{\psicin}{\psic^{\mathrm{in}}}
\def\prfe{\hspace*{\fill} $\Box$

\smallskip \noindent}

\newtheorem{Theorem}{Theorem}
\newtheorem{Proposition}{Proposition}
\newtheorem{Lemma}{Lemma}
\newtheorem{Definition}{Definition}
\newtheorem{Corollary}{Corollary}
\newtheorem{Remark}{Remark}

\title{A mathematical theory of isolated systems in relativistic plasma physics}
\author{Simone Calogero\footnote{E-Mail: calogero@mct.uminho.pt}\\[0.5cm]
Departamento de Matem\'atica para a Ci\^encia e Tecnologia\\
Campus de Azur\'em da Universidade do Minho\\
4800-058 Guimar\~aes, Portugal\\}
\date { }
\begin{document}
\maketitle
\begin{abstract}
The existence and the properties of isolated solutions to the relativistic Vlasov-Maxwell system with initial data on the backward hyperboloid $t=-\sqrt{1+|x|^2}$ are investigated. Isolated solutions of Vlasov-Maxwell can be defined by the condition that the particle density is compactly supported on the initial hyperboloid and by imposing the absence of incoming radiation on the electromagnetic field.  
Various consequences of the mass-energy conservation laws are derived by assuming the existence of smooth isolated solutions which match the inital data. In particular, it is shown that the mass-energy of isolated solutions on the backward hyperboloids and on the surfaces of constant proper time are preserved and equal, while the mass-energy on the forward hyperboloids is non-increasing and uniformly bounded by the mass-energy on the initial hyperboloid. 
Moreover the global existence and uniqueness of classical solutions in the future of the initial surface is established for the one dimensional version of the system. 
\end{abstract}

\section{Introduction}\label{Introduction}
\setcounter{equation}{0} 
Models representing
isolated systems have an important role in physics. Even though no system can be rigorously considered
isolated, it is common in physics to
disregard in first approximation the effects that external agents,
including the observer, have on the dynamics of the system under
consideration. What exactly means that a physical system is isolated depends on the underlying physical theory that is used to describe the dynamics of the model. 
In Newtonian gravity, for example, a compactly supported matter distribution is referred to as being isolated if the gravitational field energy at any fixed time decays to zero at large distances from the matter. This definition implies that the gravitational potential generated by the mass distribution equals the convolution product of the mass density with the fundamental solution of the Poisson equation. However if the dynamics of the same matter distribution is described using general relativity instead of Newtonian gravity, i.e., taking into account the relativistic effects, the precise definition of isolated sistem becomes more complicated and indeed is not yet fully understood, cf. \cite{Chr}. 

In order to clarify the kind of difficulties that arise when we try to extend the concept of isolated system from pre-relativistic to relativistic physics, we study here the evolution of an isolated charge distribution in Maxwell's electrodynamics. As opposed to the electrostatic case, where the potential solves the Poisson equation and decays at the same rate in all directions, the finite, constant (in vacuum) speed of propagation of signals implies that a typical solution of the Maxwell equations will in general allow for a net
flux of field energy coming from infinity and propagating along the null directions. The part of the field responsible for this energy flux is called incoming radiation. If the system is to be thought of as isolated, the incoming radiation should be ruled out by appropriate boundary conditions. The precise form and relevance of these conditions will be discussed later. 

As regard to the matter model, we postulate a collisionless plasma described by the relativistic Vlasov equation, which coupled to the Maxwell equations yields the relativistic Vlasov-Maxwell system. The reason for this choice is that solutions of the relativistic Vlasov-Maxwell system are in general more regular than for other matter models---such as perfect fluids---and this considerably simplifies the analysis in the sequel. In particular, the compact support property of the Vlasov matter field holds for all times provided it is required initially.
  
\subsection{The Vlasov-Maxwell system}
\setcounter{equation}{0} Consider a large ensemble of charged particles (plasma) which interact only by means of the electromagnetic forces that they generate collectively; external forces and collisions between the particles are assumed to be negligible. In the case of one species of particle, the average dynamics of the plasma is described through the particles distribution function $f=f(t,x,p)$, which satisfies the Vlasov equation:
\begin{equation}\label{vlasovt}
\partial_{t}f+\widehat{p}\cdot\nabla_{x}f+(E+\widehat{p}\times B)
\cdot\nabla_{p}f=0.
\end{equation}
Here $(t,x)\in\R\times\R^3$ denotes a Cartesian system of coordinates in the four dimensional Minkowski space (i.e., a system of coordinates in which the Minskowski metric has the canonical diagonal form) and $p\in\R^3$ is the momentum variable. The time coordinate $t$ will be called {\it proper time}. $(E,B)=(E,B)(t,x)$ denotes the mean electromagnetic field generated by the plasma; the vector field $E+\widehat{p}\times B$ is the Lorentz force and
\begin{equation}\label{velocity}
\widehat{p}=\frac{p}{\sqrt{1+|p|^{2}}}
\end{equation}
denotes the relativistic velocity of a particle with momentum $p$. Units are chosen so that the mass and the charge of each particle, as well as the speed of light, have unit value.
The electromagnetic field $(E,B)$ is required to solve the Maxwell equations
\begin{equation}\label{maxwellt}
\left\{\begin{array}{ll}\partial_{t} E=\nabla\times B- j,& \quad \nabla\cdot E=\rho,\\
\partial_{t} B=-\nabla\times E,& \quad\nabla\cdot B=0,\end{array}\right.
\end{equation}
where the charge density $\rho$ and the current density $j$ are given by
\begin{equation}\label{sourcest}
\rho(t,x)=\int f(t,x,p)\,dp,\quad j(t,x)=\int
\widehat{p}\,f(t,x,p)\,dp.
\end{equation}
The set of equations (\ref{vlasovt})--(\ref{sourcest}) is the (relativistic) Vlasov-Maxwell system.
Sufficiently regular solutions of this sistem satisfy the continuity
equation
\begin{equation}\label{conteqt}
\partial_t\rho + \nabla\cdot j=0,
\end{equation}
as well as the energy identity
\begin{equation}\label{localcont}
\partial_te + \nabla\cdot\mathfrak{p}=0,
\end{equation}
where
\begin{eqnarray*}
e(t,x) = \int \sqrt{1+|p|^2}\,f\,dp + \frac{1}{2}|E|^2 +
\frac{1}{2}|B|^2, \quad
\mathfrak{p}(t,x) = \int p\,f\,dp +E\times
B,
\end{eqnarray*}
are the local energy and the local momentum, respectively.
Due to its relativistic character, the Vlasov-Maxwell system allows for the presence of radiation fields. The radiation is defined as the part of the electromagnetic
field which carries energy to null infinity. It is distinguished in \textit{outgoing} radiation, which propagates energy to  future null
infinity, and \textit{incoming} radiation, which propagates energy to past null infinity. 
As for any solution $(f(t,x,p),E(t,x),B(t,x))$ of Vlasov-Maxwell, the triple $(f(-t,x,-p),
E(-t,x), -B(-t,x))$ gives a new solution thereof (time reflection symmetry), solutions of this system
contain in general outgoing as well as incoming radiation. Our main interest is in the solutions of the Vlasov-Maxwell system which describe an isolated plasma. The precise definition of this class of solutions will be given in Section \ref{generalproperties}. The physical idea behind the concept of isolated solution is that the system is not hit by energy coming from infinity. This condition can be achieved by requiring that (i) the particle density is supported away from past null infinity and (ii) there is no contribution from the incoming radiation in the electromagnetic field ({\it no-incoming radiation condition}). Note that since radiation propagates energy-momentum at the speed of light, the correct set of infinity where to impose the condition for having an isolated system is past null infinity.

\subsection{The initial value problem with data on a backward hyperboloid}
As the definition of isolated systems is related to the behavior of the solution as $t\to -\infty$, the standard framework of the Cauchy problem with data at $t=0$ does not seem to provide a suitable mathematical setting for the analysis of such solutions (local isolated solutions of the Cauchy problem are meaningless). In the framework of the initial value problem, it is more natural to assign the initial data of isolated solutions on a surface which cuts past null infinity. In \cite{C3} this surface was chosen to be the past light cone with vertex in the origin of the Cartesian system of coordinates. This choice leads naturally to consider the foliation of Minkowski space by the family of past light cones $t-|x|=v$ and to study the Vlasov-Maxwell system in the advanced time coordinate $v$. However this situation is complicated by the fact that, since the coefficients
of the Vlasov equation are discontinuous at the vertex of the cones, smooth solutions of the Vlasov-Maxwell system in the advanced time coordinate do not exist in general. Considering weak solutions to the problem does not seem completely satisfactory, because the singularity at the vertex of the cones is only an artefact due to a bad choice of coordinates and not a ``real" singularity.

Motivated by the previous remarks, we continue the analysis on isolated solutions of Vlasov-Maxwell started in \cite{C3} by employing a different approach, namely by studying the initial value problem with data on the backward hyperboloid $t=-\sqrt{1+|x|^2}$. Besides solving the problem on the regularity of isolated solutions mentioned above, the present paper also contains several results which were left open in \cite{C3} or merely conjectured therein.

%Let us start by rewriting the Vlasov-Maxwell system in a time coordinate adapted to the foliation of Minkowski space by backward hyperboloids.. 
Let $(\fc,\Ec,\Bc)$ denote the restriction of the solutions of Vlasov-Maxwell on the backward hyperboloids $t=\tau-\sqrt{1+|x|^2}$, $\tau\geqslant 0$. In order to make transparent the relation between $(\fc,\Ec,\Bc)$ and $(f,E,B)$, the coordinates on the hyperboloids will be chosen by orthogonal projection on the hypersurfaces of constant proper time. Hence 
\[
\fc(\tau,x,p)=f(\tau-\sqrt{1+|x|^2},x,p),\quad (\Ec,\Bc)(\tau,x)=(E,B)(\tau-\sqrt{1+|x|^2},x).
\]   
In terms of these new dynamical variables, the Vlasov-Maxwell system takes the form
\begin{equation}\label{vlasov2}
(1+\widehat{p}\cdot\widehat{x})\partial_\tau\fc+\widehat{p}\cdot\nabla_x\fc+(\Ec+\widehat{p}\times\Bc)\cdot\nabla_p\fc=0,
\end{equation}
\begin{eqnarray}
&&\partial_\tau(\Ec-\widehat{x}\times\Bc)=\nabla\times\Bc-\jc,\label{maxwell1}\\
&&\partial_\tau(\Bc+\widehat{x}\times\Ec)=-\nabla\times\Ec,\label{maxwell2}
\end{eqnarray}
\begin{eqnarray}
&&\partial_\tau(\Ec\cdot \widehat{x})+\nabla\cdot\Ec=\rhoc,\label{constrainteq1}\\
&&\partial_\tau(\Bc\cdot \widehat{x})+\nabla\cdot\Bc=0,\label{constrainteq2}
\end{eqnarray}
where
\begin{equation}\label{sourcedef}
\rhoc(\tau,x)=\int
\fc\,dp,\quad\jc(\tau,x)=\int\widehat{p}\,\fc\,dp,\quad\widehat{x}=\frac{x}{\sqrt{1+|x|^2}}.
\end{equation}
Initial data are given at $\tau=0$ and denoted by
\[
\fcin(x,p)=\fc(0,x,p),\quad\Ecin(x)=\Ec(0,x),\quad \Bcin(x)=\Bc(0,x).
\]
Later we shall discuss the equivalence of the system above
with the evolution equations (\ref{vlasov2})--(\ref{maxwell2}) and a set of constraint equations on the initial data. 
We are interested in the question of existence and
uniqueness of classical solutions in the future, i.e., for $\tau\in
[0,\infty)$, which match the initial data at $\tau=0$. These solutions will be referred to as ``global", 
although they are defined only for $\tau\geqslant 0$ (note that (\ref{vlasov2})--(\ref{sourcedef}) is not time symmetric). Obviously, data on a backward hyperboloid are in general not sufficient to determine a unique solution in the future, 
since the intersection
between the initial surface and the domain of dependence of the
solutions on a space-time point sufficiently far in the future is not a compact
set (precisely, uniqueness is lost in the region of Minkowski space in the future of the past light cone $t=-|x|$). Based on causality arguments, it is expected that a unique solution of (\ref{vlasov2})--(\ref{sourcedef}) is obtained by supplying the initial conditions with data at past null infinity (i.e., at $|x|\to +\infty$ on the surfaces $\tau=const.$). That this is the case will be proved in Section \ref{1d} for the one dimensional version of the system. In the class of solutions which admit a continuous extension at past null infinity, the no-incoming radiation condition is equivalent to giving zero data at past null infinity.

\subsection{Outline of the paper and outlook}
The outline of the present article is as follows.
In Section \ref{generalproperties}
various consequences of the local conservation laws for (\ref{vlasov2})--(\ref{sourcedef}) and of the no-incoming radiation 
condition are collected. The results of Section
\ref{generalproperties} are conditional, as they assume the existence of global
classical solutions, and improve those of \cite{C3} in several aspects. In particular, the boundedness of the mass-energy on the forward hyperboloids, which was left open in \cite{C3}, is established here by integrating the identities (\ref{conteqt})-(\ref{localcont}) in the interior of suitable space-time regions. Moreover we prove that the energy of isolated solutions on the surfaces of constant proper time is preserved and equals the energy on the initial hyperboloid, a property which was only conjectured in \cite{C3}. 
In Section \ref{sphsym} we state a global existence and
uniqueness theorem for spherically symmetric solutions. The proof is based on classical arguments introduced in \cite{Ba, GSh} and is therefore omitted.  Note that in spherical symmetry
the magnetic field vanishes identically (if decay at infinity is
imposed) and the Maxwell equations reduce to the Poisson equation
for the electric field. Hence there is neither incoming nor
outgoing radiation in spherical symmetry. In order to introduce radiation effects, we consider in Section \ref{1d} the Vlasov-Maxwell system in one spatial dimension and two dimensions in velocity (the so-called ``one and one-half" dimensional system). In this case we are able to prove the existence and uniqueness of a global classical solution with given data at $\tau=0$ and data at past null infinity. Moreover we show that if and only if the data at past null infinity vanish, the solution satisfies the no-incoming radiation condition.

To conclude the Introduction, we review some important results on the Vlasov-Maxwell system. Existence and uniqueness of local solutions to the Cauchy problem has been proved in \cite{W}. 
In \cite{GS2} it is shown that a classical solution exists in any interval of time in which the momentum support of the particles density is bounded.
Different proofs of this result are given \cite{BGP} and \cite{KS}. In \cite{Pa} it is proved that such a continuation principle of local solutions hold for a wide variety of linear equations coupled (in a non-linear way) to the Vlasov equation. 
The result in \cite{GS2} was applied to prove global existence and uniqueness under different smallness assumptions on the initial data (cf. \cite{GSh1,GS4,R}) and for arbitrarily large data in one and two space dimensions \cite{GSh2,GSh3}. Existence, but not uniqueness, of global weak solutions in three spatial dimensions is proved in \cite{DipLio}. 

Isolated solutions of the Vlasov-Maxwell system were first studied in \cite{C,C2} using the retarded solution of the equations. For related results based on approximation methods see \cite{BKRR}.
 
\section{General properties of classical solutions}\label{generalproperties}
\setcounter{equation}{0} 
In this section we study
several properties of global solutions of (\ref{vlasov2})--(\ref{sourcedef}) satisfying the regularity
condition
\begin{equation}\label{regularity}
\fc\in C^1([0,\infty)\times\R^3\times\R^3),\quad\Ec,\Bc\in
C^1([0,\infty)\times\R^3)
\end{equation}
and so they are solutions of (\ref{vlasov2})--(\ref{sourcedef}) in
a classical sense. It is also assumed that $\fc$ has bounded support
in the momentum, precisely
\begin{equation}\label{boundP}
\mathcal{P}_\cap(\tau)=\sup\{\sqrt{1+|p|^2}:\fc(s,x,p)\neq 0,\,0\leqslant
s\leqslant \tau,\,x\in\R^3\}<\infty,\quad\forall\,\tau\in [0,\infty).
\end{equation}
In particular, all the integrals in the momentum variable in the
sequel are understood to be extended over a (time dependent) compact set. For the 
sake of reference we define
\[
\mathcal{C}_\infty=\{(\fc,\Ec,\Bc) \textnormal{ which satisfy (\ref{regularity})-(\ref{boundP})}\}. 
\]
The precise mathematical definition of isolated solution of the Vlasov-Maxwell system can now be given. We denote by $S_r=\{x\in\R^3:|x|=r\}$ the sphere with center in the origin and radius $r>0$, by $dS_r$ the invariant volume element thereon and by $k=x/|x|$ the outward unit normal to the sphere.
\begin{Definition}\label{isolatedsolutions}
A solution $(\fc,\Ec,\Bc)\in\mathcal{C}_\infty$ of (\ref{vlasov2})--(\ref{sourcedef}) is said to be \textnormal{isolated} if 
\begin{itemize}
\item[(A)] There exists a constant $R_0>0$ such that 
\[
\rhoc(\tau,x)=0,\quad \textnormal{{\it for}}\ \sqrt{1+|x|^2}\geqslant R_0+\frac{1}{2}\tau;
\] 
\item[(B)] The no-incoming radiation condition (NIRC) in the future is satisfied, namely
\begin{equation}\label{NIRC}
\lim_{r\to\infty}\int_{\tau_1}^{\tau_2}\int_{S_r}k\cdot\left[\Ec\times\Bc\right](\tau,x)\,dS_r\,d\tau=0,\quad\forall\,\tau_1,\,\tau_2\geqslant 0.
\end{equation}
\end{itemize}
\end{Definition}
\begin{Remark}\textnormal{A local isolated solution in the interval $[0,\mathcal{T})$, $\mathcal{T}>0$, can be defined by imposing
the conditions (A) and (B) only for $\tau,\tau_1,\tau_2\in[0,\mathcal{T})$.}
\end{Remark} 
\begin{Remark}\textnormal{The limit in the left hand side of (\ref{NIRC}) defines the energy carried out by the electromagnetic field to past null infinity in the interval $[\tau_1,\tau_2]$ of the time coordinate $\tau$. As we are interested only in going forward in time, we require the NIRC only in the future of the initial hyperboloid $\tau=0$.}
\end{Remark}
\begin{Remark}\textnormal{The condition (A) means geometrically that the matter is supported away from the region of past null infinity in the future of the initial hyperboloid (however the matter support may intersect future null infinity). Although this assumption could be relaxed and replaced by a decay condition that implies the absence of kinetic energy at past null infinity, we shall avoid to do so, as we feel that such a condition does not add any phyisically interesting new feature to the concept of isolated system.}
\end{Remark} 
It turns out that the condition (A) in Definition \ref{isolatedsolutions} is satisfied by any solution in the class $\mathcal{C}_\infty$ if the initial datum $\fcin$ is chosen with compact support in $x$ and defining 
\begin{equation}\label{R0}
R_0=\inf\{r>0:\fcin(x,p)=0,\,\sqrt{1+|x|^2}\geqslant r,\,p\in\R^3\},
\end{equation}
see Lemma \ref{supportf}. 
For this reason, the initial data are throughout assumed to satisfy
\begin{equation}\label{initialdata}
0\leqslant\fcin\in C^1_c(\R^3\times\R^3),\quad\Ecin,\Bcin\in
C^2(\R^3).
\end{equation}
At this point it is worth to emphasize that even if the particles density were given with compact support on the surface $t=0$ (as it is usually the case when studying the Cauchy problem), there is no guarantee {\it a priori} that it will have compact support also on the backward hyperboloid $t=-\sqrt{1+|x|^2}$. This already indicates that the class of solutions of (\ref{vlasov2})--(\ref{sourcedef}) with data as in (\ref{initialdata}) is strictly contained in the class of solutions with data prescribed at $t=0$.
\begin{Remark}\textnormal{Definition \ref{isolatedsolutions} can be adopted also for other matter models coupled to the Maxwell system, such as perfect fluids described by the Euler equations. However in the latter case it is not enough to give a charge density with compact support in the initial hyperboloid to make the condition (A) satisfied. To this purpose one needs to formulate an initial-(free) boundary value problem for the Maxwell-Euler equations, for which there is no existence result available.}
\end{Remark}
We split our analysis in six different subsections. The results of Sections \ref{vlasovequation} and \ref{sphsym} make use explicitely of the Vlasov equation, whereas the content of Sections \ref{maxwellequations}--\ref{unique} applies to any matter model for which the local conservation laws of mass and energy hold.

\subsection{The Vlasov equation}\label{vlasovequation}
Let us start by pointing out some basic properties of $\fc$. The estimate
\begin{eqnarray*}
1+\widehat{p}\cdot\widehat{x}&=&1+\frac{p\cdot
x}{\sqrt{1+|p|^2}\sqrt{1+|x|^2}}\geqslant 1-\frac{|p||x|}{\sqrt{1+|p|^2}\sqrt{1+|x|^2}}\nonumber\\
&\geqslant&1-\frac{|p|}{\sqrt{1+|p|^2}}=\frac{1}{\sqrt{1+|p|^2}(\sqrt{1+|p|^2}+|p|)}\geqslant
\frac{1}{2(1+|p|^2)}\label{est}
\end{eqnarray*}
shows that for solutions in the class $\mathcal{C}_\infty$, the equation
(\ref{vlasov2}) is equivalent to
\begin{equation}
\partial_\tau\fc+\frac{p}{p_0}\cdot\nabla_x\fc+\frac{1}{p_0}\left(\sqrt{1+|p|^2}\Ec+p\times\Bc\right)\cdot\nabla_p\fc=0,\label{vlasov3}
\end{equation}
where $p_0$ is defined by
\[
p_0=\sqrt{1+|p|^2}+p\cdot\widehat{x}  > 0.
\]

The characteristics of the differential operator in the left hand
side of (\ref{vlasov3}) are the solutions of
\begin{equation}\label{char}
\dot{x}=\frac{p}{p_0},\quad
\dot{p}=\frac{1}{p_0}\left(\sqrt{1+|p|^2}\Ec+p\times\Bc\right)
\end{equation}
and we denote by $(X,P)(s)$ the
characteristic curve satisfying $(X,P)(\tau)=(x,p)$; note that $(X,P)$ also depends on $(\tau,x,p)$, but this is not reflected in our notation. Let $F=(F_1,F_2)$ denote the right hand side of (\ref{char}), that is
\begin{equation}\label{force}
F_1(x,p)=\frac{p}{p_0},\quad F_2(x,p)=\frac{1}{p_0}\left(\sqrt{1+|p|^2}\Ec+p\times\Bc\right).
\end{equation}
Since $F$ is a $C^1$ function when $(\Ec,\Bc)\in C^1$, then the characteristics are smooth and one can write the solution of (\ref{vlasov3}) as
\begin{equation}\label{reprf}
\fc(\tau,x,p)=\fcin((X,P)(0)).
\end{equation}
In particular, $\fc$ remains non-negative for all times and $\|\fc(\tau)\|_\infty\leqslant\|\fcin\|_\infty$.

We shall now derive the conservation laws satisfied by the
solutions of (\ref{vlasov3}). A straightforward computation
shows that the vector (\ref{force})
satisfies
\begin{equation}\label{identity}
\left[\nabla_{(x,p)}\cdot F\right]\left(s,X(s),P(s)\right)
=-\frac{d}{ds}\log\left(1+\widehat{P}(s)\cdot \widehat{X}(s)\right),
\end{equation}
where $\widehat{P}=P/\sqrt{1+|P|^2}$. In fact, each side of
(\ref{identity}) equals, along characteristics,
\[
-\frac{1}{(1+\widehat{p}\cdot \widehat{x})^2}\left [\frac{|\widehat{p}|^2-
(\widehat{p}\cdot\widehat{x})^2}{\sqrt{1+|x|^2}}+\frac{1}{\sqrt{1+|p|^2}}\Big(\Ec\cdot
(\widehat{x}-(\widehat{p}\cdot \widehat{x})\widehat{p})-(\widehat{p}\times
\widehat{x})\cdot\Bc\Big)\right].
\]
From (\ref{identity}) we deduce
\[
\det\left[\frac{\partial(X,P)(s)}{\partial(x,p)}\right]=\frac{1+\widehat{p}\cdot
\widehat{x}}{1+\widehat{P}(s)\cdot \widehat{X}(s)}.
\]
Hence using (\ref{reprf}) the next lemma follows.
\begin{Lemma}
For any solution of (\ref{vlasov2})--(\ref{sourcedef}) in the class $\mathcal{C}_\infty$ and for any measurable function $Q:\R\to\R$, we have
\[
\int\int Q(\fc)(1+\widehat{p}\cdot \widehat{x})\,dp\,dx=const.
\]
In particular, by choosing $Q(z)=z^q$, $q\geqslant 1$,
\begin{equation}\label{lqest}
\|(1+\widehat{p}\cdot
\widehat{x})^{1/q}\fc(\tau)\|_{L^q(\R^3\times\R^3)}=const.
\end{equation}
\end{Lemma}

In the next lemma we
estimate the $x-$support of $\fc$.
\begin{Lemma}\label{supportf}
Let $(\fc,\Ec,\Bc)\in\mathcal{C}_\infty$ be a solution of (\ref{vlasov2})--(\ref{sourcedef}) with initial data (\ref{initialdata}). Then 
\[
\fc(\tau,x,p)=0,\ \textnormal{ {\it for} } \sqrt{1+|x|^2}\geqslant R_0+\frac{1}{2}\tau,\  \tau\geqslant 0,
\]
where $R_0>0$ is given by (\ref{R0}).
\end{Lemma}
\noindent\textit{Proof: } For all $0\leqslant s\leqslant \tau$ we have, by the
first equation in (\ref{char}),
\begin{eqnarray*}
\sqrt{1+|x|^2}&=&\sqrt{1+|X(s)|^2}+\int_s^\tau\frac{P(\tau')\cdot
\widehat{X}(\tau')}{\sqrt{1+|P(\tau')|^2}+P(\tau')\cdot \widehat{X}(\tau')}d\tau'\\
&\leqslant& \sqrt{1+|X(s)|^2}+\int_{\mathcal{I}^+}\frac{P(\tau')\cdot
\widehat{X}(\tau')}{\sqrt{1+|P(\tau')|^2}+P(\tau')\cdot \widehat{X}(\tau')}d\tau',
\end{eqnarray*}
where $\widehat{X}=X/\sqrt{1+|X|^2}$ and $
\mathcal{I}^+=\{\tau'\in [0,\tau]:(P\cdot \widehat{X})(\tau')\geqslant 0\}$.
Since $\sqrt{1+|p|^2}>|p\cdot \widehat{x}|$ we obtain
\begin{eqnarray*}
\sqrt{1+|x|^2}&\leqslant& \sqrt{1+|X(s)|^2}+\frac{1}{2}(\tau-s).
\end{eqnarray*}
For $s=0$ this implies $\sqrt{1+|x|^2}\leqslant R_0+\frac{1}{2}\tau$ in the support of $\fc$. \prfe
\begin{Corollary}
A solution $(\fc,\Ec,\Bc)\in\mathcal{C}_\infty$ with initial data (\ref{initialdata}) is isolated if and only if it satisfies the NIRC in the future.
\end{Corollary}

\subsection{The Maxwell equations}\label{maxwellequations}
Next the Maxwell equations (\ref{maxwell1})--(\ref{constrainteq2}) will be considered. A basic fact is that they are equivalent to a set of constraint equations on the initial data and a set of evolution equations. As opposed to the case of initial data on a past light cone considered in \cite{C3}, data on a backward hyperboloid are restricted by the same number of constraints as for the case of data given on the surface $t=0$ (in \cite{C3} there appear additional constraints because the initial data are given on a characteristic surface). We consider the system (\ref{maxwell1})--(\ref{constrainteq2}) for a given pair $(\rhoc,\jc)\in C^1$ satisfying the equation
\begin{equation}\label{conteq2}
\partial_\tau(\rhoc+\jc\cdot\widehat{x})=-\nabla\cdot\jc.
\end{equation}
Scalar multiplying (\ref{maxwell1})-(\ref{maxwell2}) by $\widehat{x}$ and using (\ref{constrainteq1})-(\ref{constrainteq2}) we obtain
\begin{equation}\label{constraintequations}
\nabla\cdot\Ec=(\rhoc+\jc\cdot\widehat{x})-\widehat{x}\cdot\nabla\times\Bc,\quad\nabla\cdot\Bc=\widehat{x}\cdot\nabla\times\Ec.
\end{equation}
A simple vector algebra computation shows that the equations (\ref{constraintequations}) are satisfied for all times provided they hold at $\tau=0$ and the functions $\Ec,\Bc,\rhoc,\jc$ satisfy
(\ref{maxwell1}), (\ref{maxwell2}) and (\ref{conteq2}). Hence we have the following
\begin{Proposition}\label{equivalence}
Let $(\rhoc,\jc)\in C^1$ be given which satisfy (\ref{conteq2}). Then
$(\Ec,\Bc)$ is a solution of (\ref{maxwell1})--(\ref{constrainteq2}) with initial data $(\Ecin,\Bcin)$ if and only if it is a solution of (\ref{maxwell1})--(\ref{maxwell2}) with initial data satisfying the \textnormal{constraint equations}
\begin{equation}\label{constrequations2}
\nabla\cdot\Ecin=(\rhoc^{\mathrm{in}}+\jc^{\mathrm{in}}\cdot\widehat{x})-\widehat{x}\cdot\nabla\times\Bcin,\quad\nabla\cdot\Bcin=\widehat{x}\cdot\nabla\times\Ecin,
\end{equation}
where $(\rhoc^{\mathrm{in}},\jc^{\mathrm{in}})=(\rhoc,\jc)_{|\tau=0}$. Moreover the solution satisfies (\ref{constraintequations}) for all $\tau\geqslant 0$.
\end{Proposition}
Proposition \ref{equivalence} applies to solutions of (\ref{vlasov2})--(\ref{sourcedef}) with
$\rhoc^{\mathrm{in}}=\int\fcin\,dp$ and $\jc^{\mathrm{in}}=\int\widehat{p}\fcin\,dp$, since the validity of (\ref{conteq2}) follows by the Vlasov equation (or by a simple change of variable in (\ref{conteqt})).

\subsection{Evolution of the mass functions}\label{evolutionmass}
Our next purpose is to study the evolution of certain mass functions associated to a solution in the class $\mathcal{C}_\infty$.
For $\tau\geqslant 0$ fixed and $\delta=0,1,2$, consider the hypersurfaces in Minkowski space which, in the Cartesian coordinates $(t,x)$, are defined by
\[
\Sigma_\delta(\tau)=\{(t,x):t+(1-\delta)\sqrt{1+|x|^2}=\tau,\, x\in\R^3\}.
\]
It follows that $\Sigma_\delta(\tau)$ is a backward hyperboloid for $\delta=0$, a surface of constant proper time for $\delta=1$ and a forward hyperboloid for $\delta=2$. The notation $[h]_{\Sigma_\delta(\tau)}$ will be used to indicate that the function $h=h(t,x)$ has to be evaluated on the surface $\Sigma_\delta(\tau)$. For $\tau\geqslant 0$ and $r>0$ fixed we set
\[
n_\delta(\tau,r)=\int_{B_r}[\rho+(1-\delta) j\cdot\widehat{x}]_{\Sigma_\delta(\tau)}\,dx,
\]
where $B_r$ denotes the ball with center in the origin and radius $r>0$. In terms of a solution $(\fc,\Ec,\Bc)\in\mathcal{C}_\infty$ of (\ref{vlasov2})--(\ref{sourcedef}), the functions $n_\delta$ are given by
\[
n_\delta (\tau,r)=\int_{B_r}\left[\rhoc+(1-\delta) \jc\cdot\widehat{x}\right]\left(\tau+\delta\sqrt{1+|x|^2},x\right)\,dx.
\]
We set 
\[
N_\delta(\tau)=\lim_{r\to\infty}n_\delta(\tau,r),
\]
which is the total mass of particles on the surface $\Sigma_\delta(\tau)$. The limit in the previous definition exists, since $n_\delta(\tau,\cdot)$ is non-decreasing. It is clear that $N_0$ can be defined also for local solutions, whereas $N_1$ and $N_2$ make sense only for global solutions. Let $N_0^{\mathrm{in}}=N_0(0)$,
which depends only on the given initial datum $\fcin$ and is bounded for data as in (\ref{initialdata}).
 
\begin{Proposition}\label{evolutionN}
Let $(\fc,\Ec,\Bc)\in\mathcal{C}_\infty$ be a solution of (\ref{vlasov2})--(\ref{sourcedef}) with initial data (\ref{initialdata}). Then the following holds:
\begin{itemize}
\item[(1)] $N_1(\tau)=N_0(\tau)=N_0^{\mathrm{in}}$, for all $\tau\geqslant 0$;
\item[(2)] $N_2(\tau_2)\leqslant N_2(\tau_1)\leqslant N_0^{\mathrm{in}}$, for all $0\leqslant\tau_1\leqslant\tau_2$;
\item[(3)] If $\sup\{\mathcal{P}_\cap(\tau),\,\tau\in [0,\infty)\}<\infty$, then $N_2(\tau)=N_0^{\mathrm{in}}$, for all $\tau\geqslant 0$.
\end{itemize}
\end{Proposition}
%\end{itemize}
%\end{Theorem}
\noindent\textit{Proof: }The equality $N_0(\tau)=N_0^{\mathrm{in}}$ corresponds to (\ref{lqest}) for $q=1$. Integrating the continuity equation (\ref{conteqt}) in the interior of the space-time region 
\[
\mathcal{R}_1(\tau,r)=\{(t,x):\tau-\sqrt{1+|x|^2}\leqslant t\leqslant\tau,\,x\in B_r\}
\]
and applying the divergence theorem we obtain
\begin{equation}\label{crucialformula}
n_1(\tau,r)=n_0(\tau,r)-\int_\tau^{\tau+\sqrt{1+r^2}}\int_{ S_r}[\jc\cdot k](\tau',x)\,dS_r\,d\tau'.
\end{equation}
By Lemma \ref{supportf} and (\ref{crucialformula}), 
\[
n_1(\tau,r)=n_0(\tau,r),\quad \textnormal{ for } r>\left[\left(2R_0+\tau\right)^2-1\right]^{1/2}, 
\]
which in the limit $r\to\infty$ implies $N_1(\tau)=N_0(\tau)$. By (\ref{conteqt}), the function $n_2$ satisfies the equation
\begin{equation}\label{n2evol}
\left(\partial_\tau-\frac{\sqrt{1+r^2}}{r}\partial_r\right)n_2=-\frac{\sqrt{1+r^2}}{r}\int_{S_r}[\rho]_{\Sigma_2(\tau)}\,dS_r.
\end{equation}
Integrating along the characteristics of the operator in the left hand side of (\ref{n2evol}) we obtain 
\[
n_2\left(\tau_2,\left[(\sqrt{1+r^2}-\tau_2)^2-1\right]^{1/2}\right)\leqslant n_2\left(\tau_1,\left[(\sqrt{1+r^2}-\tau_1)^2-1\right]^{1/2}\right),
\]
for $\tau_2\geqslant \tau_1$ and $r\geqslant\sqrt{(1+\tau_2)^2-1}$.
Letting $r\to\infty$ in the previous inequality shows that $N_2(\tau)$ is non-increasing. To complete the proof of (2) it remains to show that $N_2(0)\leqslant N_0^\mathrm{in}$. To this purpose we integrate (\ref{conteqt}) in the interior of 
\[
\mathcal{R}_2(r)=\{(t,x):0\leqslant t\leqslant \sqrt{1+|x|^2},\,x\in B_{r-t}\},\quad r>1,
\]
and apply again the divergence theorem to obtain the identity
\[
n_2\left(0,\frac{r^2-1}{2r}\right)=n_1(0,r)-\int_0^{\frac{1+r^2}{2r}}\int_{S_{r-t}}\left[\rho+j\cdot k\right](t,x)\,dS_{r-t}\,dt.
\]
Since the integral in the right hand side is positive, we obtain the inequality 
\[
n_2\left(0,\frac{r^2-1}{2r}\right)\leqslant n_1(0,r),
\]
whence, in the limit $r\to \infty$, $N_2(0)\leqslant N_1(0)=N_0^{\mathrm{in}}$. To prove (3) we use the identity
\begin{equation}\label{temporale}
n_2(\tau,r)=n_1(\tau,r)-\int_{\tau+\sqrt{1+r^2}}^{\tau+2\sqrt{1+r^2}}\int_{S_r}[\jc\cdot k](\tau',x)\,\,dS_r\,d\tau',
\end{equation}
which follows integrating (\ref{conteqt}) in the interior of the space-time region 
\[
\mathcal{R}_3(\tau,r)=\{(t,x):\tau\leqslant t\leqslant \tau+\sqrt{1+|x|^2},\,x\in B_r\}.
\]
Let $D=\sup\{\mathcal{P}_\cap(\tau),\,\tau\in [0,\infty)\}$ and $x\in\R^3:\,\fc(\tau,x,p)\neq 0$ for some $p\in\R^3$. The assumption $D<\infty$ implies 
\[
\sqrt{1+|p|^2}\geqslant\frac{\sqrt{1+D^2}}{D}|p|\geqslant
\frac{\sqrt{1+D^2}}{D} (p\cdot \widehat{x})
\]
in the support of $\fc$ and so, as in the proof of Lemma
\ref{supportf},
\begin{eqnarray*}
\sqrt{1+|x|^2}&\leqslant& \sqrt{1+|X(0)|^2}+\int_{\mathcal{I}^+}\frac{P(\tau')\cdot\widehat{X}(\tau')}{\sqrt{1+|P(\tau')|^2}+P(\tau')\cdot \widehat{X}(\tau')}d\tau'\\
&\leqslant&
R_0+\frac{D}{D+\sqrt{1+D^2}}\,\tau,
\end{eqnarray*}
for all $(x,p)\in {\rm supp}\,\fc(\tau)$, where
$\mathcal{I}^+=\{\tau'\in [0,\tau]:(P\cdot \widehat{X})(\tau')> 0\}$. This
implies that 
\[
\fc(\tau,x,p)=0,\textnormal{ for } \sqrt{1+|x|^2}\geqslant R_0+\frac{1}{2}\tau,\quad a=\frac{D}{D+\sqrt{1+D^2}}\in\left[0,\frac{1}{2}\right).
\]%\end{equation}
Whence
\[
\int_{\tau+\sqrt{1+r^2}}^{\tau+2\sqrt{1+r^2}}\int_{S_r}\left[\jc\cdot k\right](\tau',x)\,\,dS_r\,d\tau'=0\textnormal{ for } r>\left[\left(\frac{R_0+a\tau}{1-2a}\right)^2-1\right]^{1/2}.
\]
Thus (3) follows letting $r\to\infty$ in (\ref{temporale}). \prfe

If $N_2(\tau_2)<N_2(\tau_1)$,  for $0\leqslant \tau_1< \tau_2$, the difference
$N_2(\tau_1)-N_2(\tau_2)$ defines the mass (or number) of particles lost at future null
infinity in the interval $[\tau_1,\tau_2]$. This {\it outgoing mass} is measured by observers at future null infinity. Under the assumption in (3) of Proposition \ref{evolutionN}, no particles reach future null infinity and therefore the outgoing mass is zero. We remark that a uniform bound on $\mathcal{P}_\cap(\tau)$ seems to hold only under severe assumptions, such us spherical symmetry (see Section \ref{sphsym}), or small data, see \cite{C,GS4,R}.
\begin{Remark}
\textnormal{The conclusions of Proposition \ref{evolutionN} are valid even if the solution does not satisfy the NIRC.}
\end{Remark}

\subsection{Evolution of the energy functions}\label{evolutionenergy}
Now we define the energy of a solution of the Vlasov-Maxwell system on the hypersurface $\Sigma_\delta(\tau)$ as 
\[
M_\delta(\tau)=\lim_{r\to\infty} m_\delta(\tau,r),\quad m_\delta(\tau,r)=\int_{B_r}[e+(1-\delta)\mathfrak{p}\cdot\widehat{x}]_{\Sigma_\delta(\tau)}\,dx,
\]
and we set $M_0^{\mathrm{in}}=M_0(0)$.
The next goal is to study the evolution of these energy functions. We proceed along the same lines as for the evolution of the mass functions.
\begin{Proposition}\label{evolutionM}
Let $(\fc,\Ec,\Bc)\in\mathcal{C}_\infty$ be an isolated solution of (\ref{vlasov2})--(\ref{sourcedef}) (or, equivalentely, let the solution satisfy the NIRC in the future and the initial conditions (\ref{initialdata})) and assume $M_0^{\mathrm{in}}<\infty$. Then the following holds:
\begin{itemize}
\item[(1)] $M_1(\tau)=M_0(\tau)= M_0^{\mathrm{in}}$, for all $\tau\geqslant 0$;
\item[(2)] $M_2(\tau_2)\leqslant M_2(\tau_1)\leqslant M_0^{\mathrm{in}}$, for all $0\leqslant\tau_1\leqslant\tau_2$.
\end{itemize}
\end{Proposition}
\noindent\textit{Proof: }By (\ref{localcont}) we have 
\[
\partial_\tau m_0=-\int_{S_r}\left[\pc\cdot k\right](\tau,x)\,dS_r.
\]
Integrating in the interval $[0,\tau]$ we obtain
\begin{equation}\label{m0}
m_0(\tau,r)=m_0(0,r)-\int_0^{\tau}\int_{S_r}\left[\pc\cdot k\right](\tau',x)\,dS_r\,d\tau'.
\end{equation}
In the limit $r\to\infty$, the integral in the right hand side converges to zero by the support property of $\fc$ and the NIRC. This shows that $M_0(\tau)$ is constant. Now, again by (\ref{localcont}),
\[
\left(\partial_\tau\pm\partial_r\right)m_1=\int_{S_r}\left[\pm e-\mathfrak{p}\cdot k\right]_{\Sigma_1(\tau)}\,dS_r.
\] 
Since the right hand side of the previous equation is positive in the plus case and negative in the minus case, it follows (integrating along characteristics and letting $r\to\infty$) that $M_1$ is constant. To complete the proof of (1), it remains to shows that $M_1(0)=M_0^{\mathrm{in}}$. Integrating the energy identity (\ref{localcont}) in the interior of the space-time region
\[
\mathcal{R}_4(r)=\{(t,x):-\sqrt{1+|x|^2}\leqslant t\leqslant 0,\,x\in B_{r+t}\},\ r>1,
\]
and applying the divergence theorem we obtain
\begin{eqnarray}
m_1(0,r)&=&m_0\left(0,\frac{r^2-1}{2r}\right)+\int_{-\frac{1+r^2}{2r}}^0\int_{S_{r+t}}\left[e-\mathfrak{p}\cdot k\right](t,x)\,dS_{r+t}\,dt\nonumber\\
&\geqslant& m_0\left(0,\frac{r^2-1}{2r}\right).\label{m1>m0}
\end{eqnarray}
On the other hand, integrating (\ref{localcont}) in the interior of the space-time region
\begin{eqnarray*}
\mathcal{R}_5(r_1,r_2)&=&\left\{(t,x):-\sqrt{1+|x|^2}\leqslant t\leqslant\min\left(0,r_1-\sqrt{1+|x|^2}\right),\,x\in B_{r_2}\right\},\\
&&r_1>1,\,r_2>\sqrt{r_1^2-1},
\end{eqnarray*}
we obtain
\begin{eqnarray*}
m_1\left(0,\sqrt{r_1^2-1}\right)&=&m_0(0,r_2)-\int_0^{r_1}\int_{S_{r_2}}\left[\pc\cdot k\right](\tau,x)\,dS_{r_2}\,d\tau\\
&&-\int_{\sqrt{r_1^2-1}\leqslant |x|\leqslant r_2}\left[\ec+\pc\cdot\widehat{x}\right](r_1,x)\,dx\\
&&\\
&\leqslant&M_0^{\mathrm{in}}-\int_0^{r_1}\int_{S_{r_2}}\left[\pc\cdot k\right](\tau,x)\,dS_{r_2}\,d\tau.
\end{eqnarray*}
As we are considering isolated solutions, the integral in the right hand side of the latter inequality converges to zero in the limit $r_2\to\infty$, which entails $m_1(0,r)\leqslant M_0^{\mathrm{in}}$, for all $r>0$. Combining with (\ref{m1>m0}) we get
\[
m_0\left(0,\frac{r^2-1}{2r}\right)\leqslant m_1(0,r)\leqslant M_0^{\mathrm{in}},\quad\forall\,r>1,
\]
whence $M_1(0)=M_0^{\mathrm{in}}$. The equation
\begin{equation}\label{m2evol}
\left(\partial_\tau-\frac{\sqrt{1+r^2}}{r}\partial_r\right)m_2=-\frac{\sqrt{1+r^2}}{r}\int_{S_r}[e]_{\Sigma_2(\tau)}\,dS_r,
\end{equation}
shows that $M_2$ is non-increasing. The estimate $M_2(0)\leqslant M_2^{\mathrm{in}}$, which concludes the proof of (2), follows by the same argument used in Proposition \ref{evolutionN} to establish the bound $N_2(0)\leqslant N_2^{\mathrm{in}}$, i.e., integrating the energy identity (\ref{localcont}) in the interior of the space-time region $\mathcal{R}_2(\tau)$.\prfe

If $M_2(\tau_2)<M_2(\tau_1)$,  for $0\leqslant \tau_1< \tau_2$, the difference
$M_2(\tau_1)-M_2(\tau_2)$ measures the energy dissipated at future null
infinity in the interval $[\tau_1,\tau_2]$. This {\it outgoing energy} is measured by observers at future null infinity. 
\begin{Remark}
\textnormal{One clearly expects that the absence of outgoing energy implies the absence of outgoing mass. To see this, note that (\ref{n2evol}) and (\ref{m2evol}) imply 
\[
\left(\partial_\tau-\frac{\sqrt{1+r^2}}{r}\partial_r\right)(m_2-n_2)=-\frac{\sqrt{1+r^2}}{r}\int_{S_r}\left([e]_{\Sigma_2(\tau)}-[\rho]_{\Sigma_2(\tau)}\right)\,dS_r\leqslant 0,
\]
which entails $M_2(\tau_2)-M_1(\tau_1)\leqslant N_2(\tau_2)-N_1(\tau_1)\leqslant 0$. Hence $M_1(\tau_1)=M_2(\tau_2)$ implies $N_1(\tau_1)=N_2(\tau_2)$.}
\end{Remark}
\begin{Remark}
\textnormal{If the solution is not isolated, then the energy on the backward hyperboloids is non-decreasing. For the function $m_0$ satisfies the equation
\[
\left(\partial_\tau+\frac{\sqrt{1+r^2}}{r}\partial_r\right)m_0=\frac{\sqrt{1+r^2}}{r}\int_{S_r}[e]_{\Sigma_0(\tau)}\,dS_r\geqslant 0,
\]
whence $M_0(\tau_2)\geqslant M_0(\tau_1)$, for all $\tau_2\geqslant\tau_1$.}
\end{Remark}

\subsection{A uniqueness theorem}\label{unique}
Let us show that the NIRC guarantees uniqueness of solutions to the Maxwell equations with data on a backward hyperboloid.
\begin{Proposition}\label{uniqueness}
Let $\rhoc,\jc\in C^1$ be given such that the continuity equation (\ref{conteq2}) is satisfied. Let $(\Ecin,\Bcin)\in C^1(\R^3)$ be given such that the constraint equations (\ref{constrequations2}) are satisfied and $M_0(0)<\infty$, where
\[
M_0(\tau)=\frac{1}{2}\int\left(|\Ec|^2+|\Bc|^2+2(\Ec\times\Bc)\cdot\widehat{x}\right)\,dx.
\]
There exists at most one solution $(\Ec,\Bc)\in C^1$ of the Maxwell system (\ref{maxwell1})--(\ref{constrainteq2}) satisfying (\ref{NIRC}) and $(\Ec,\Bc)_{|t=0}=(\Ecin,\Bcin)$.
\end{Proposition}
\noindent\textit{Proof:} The difference $(\delta\Ec,\delta\Bc)$ of two solutions with the same data satisfies (\ref{maxwell1})--(\ref{constrainteq2}) with $\rhoc=\jc=0$ and zero data; by Proposition \ref{equivalence}, $(\delta\Ec,\delta \Bc)$ solves the equations
\begin{equation}\label{tempo3}
\partial_\tau(\delta\Ec-\widehat{x}\times\delta\Bc)=\nabla\times\delta\Bc,\quad\partial_\tau(\delta\Bc+\widehat{x}\times\delta\Ec)=-\nabla\times\delta\Ec,
\end{equation}
\begin{equation}\label{tempo4}
\nabla\cdot\delta\Ec=-\widehat{x}\cdot\nabla\times\delta\Bc,\quad\nabla\cdot\delta\Bc=\widehat{x}\cdot\nabla\times\delta\Ec.
\end{equation}
On the other hand, the energy of $(\delta\Ec,\delta\Bc)$ on the backward hyperboloids is zero; this implies 
\begin{eqnarray*}
0&\leqslant& |\delta\Ec\cdot\widehat{x}|^2+|\delta\Bc\cdot\widehat{x}|^2+|\delta\Ec-\widehat{x}\times\delta\Bc|^2+|\delta\Bc+\widehat{x}\times\delta\Ec|^2\\
&=&\frac{1+2|x|^2}{1+|x|^2}\left(|\delta\Ec|^2+|\delta\Bc|^2\right)+4(\delta\Ec\times\delta\Bc)\cdot\widehat{x}\\
&\leqslant&4\left(\frac{1}{2}|\delta\Ec|^2+\frac{1}{2}|\delta\Bc|^2+(\delta\Ec\times\delta\Bc)\cdot\widehat{x}\right)=0,
\end{eqnarray*}
whence $\delta\Ec-\widehat{x}\times\delta\Bc=\delta\Bc+\widehat{x}\times\delta\Ec=0$, which together with (\ref{tempo3})-(\ref{tempo4}) yields $\delta\Ec=\delta \Bc=0$. \prfe

\subsection{Spherically symmetric solutions}\label{sphsym}
\setcounter{equation}{0} In spherical symmetry we have
$\nabla\times\Ec=\nabla\times\Bc=0$ and so, by the second equation in (\ref{constraintequations}),
$\nabla\cdot\Bc=0$. Under the additional boundary condition
$\lim_{|x|\to\infty}\Bc=0$, this implies that the magnetic field
vanishes identically. Thus all spherically symmetric solutions satisfy the NIRC and so, choosing $\fcin$ of compact support, they are all isolated. Moreover, by the first equation in (\ref{constraintequations}), 
\begin{eqnarray}
\Ec(\tau,x)&=&\frac{k}{r^2}\int_0^r(\rhoc+\jc\cdot\widehat{x})(\tau,r')r'^2\,dr'\nonumber\\
&=&\frac{1}{4\pi}\int\frac{(x-y)}{|x-y|^3}(\rhoc+\jc\cdot\widehat{x})(\tau,y)\,dy,\label{ess}
\end{eqnarray}
the second equality being valid in spherical symmetry. By abuse of notation we use the same symbol to denote a
spherically symmetric function in spherical and Cartesian
coordinates.
The Vlasov equation reduces to
\begin{equation}
\partial_\tau\fc+\frac{p}{p_0}\cdot\nabla_x\fc+\frac{\sqrt{1+|p|^2}}{p_0}\Ec\cdot\nabla_p\fc=0.\label{vlasovss}
\end{equation}
Note also the conservation of angular momentum; along characteristics,
\[
\frac{d}{ds}|x\times p|^2=0.
\]
In the spherically symmetric case we have the following global existence theorem.
\begin{Theorem}\label{global}
Let $0\leqslant \fcin\in
C^1_c(\R^3\times\R^3)$ be spherically symmetric;
there exists a unique (spherically symmetric) $\fc\in C^1([0,\infty)\times\R^3\times\R^3)$ solution  of
(\ref{ess})-(\ref{vlasovss}) such that $\fc(0,x,p)=\fcin(x,p)$.
Moreover, there exists a constant $D>0$, depending only on bounds
on the initial datum, such that
\begin{equation}\label{estP}
\mathcal{P}_\cap(\tau)\leqslant D.
\end{equation}
\end{Theorem}
The proof of Theorem \ref{global} is a straightforward adaptation of the argument
given in \cite[Theorem II]{GSh}. Since the details of this adaptation have already been given in \cite{C3},
they will not be repeated here. Note however that in \cite{C3} an extra condition 
on the initial datum is assumed, namely that there are no particles with arbitrarily small
angular momentum at time zero. By the conservation of angular momentum, this condition is 
preserved in time, and since the maximal momentum of the particles is uniformly bounded, the conclusion 
follows that the particle density vanishes in a neighbourhood of the axis $r=0$. This property was necessary
in the analysis of \cite{C3} because, as we mentioned in the Introduction, the particle density is singular on
the vertex of the backward light cones. In the case considered here, however, this problem does not arise, and no further condition
is needed on the initial datum. 
%because the characteristics of the Vlasov equation are smooth everywhere in the time 
%coordinate $\tau$, included at the axis $r=0$.
\begin{Corollary}   
For the solution of Theorem \ref{global},
\[
N_2(\tau)=N_1(\tau)=N_0(\tau)=N_0^{\mathrm{in}},
\]
\[
M_2(\tau)=M_1(\tau)=M_0(\tau)=M_0^{\mathrm{in}}.
\]
\end{Corollary}
\noindent\textit{Proof: }The equalities $N_1(\tau)=N_0(\tau)=N_0^{\mathrm{in}}$ and $M_1(\tau)=M_0(\tau)=M_0^{\mathrm{in}}$ hold for all isolated solutions, whereas $N_2(\tau)=N_0^{\mathrm{in}}$ follows by (\ref{estP}), see (3) of Proposition \ref{evolutionN}. For $B=0$, the function $m_2$ satisfies an equation similar to (\ref{temporale}), namely
\[
m_2(\tau,r)=m_1(\tau,r)-\int_{\tau+\sqrt{1+r^2}}^{\tau+2\sqrt{1+r^2}}\int_{S_r}\int p\cdot
k\,\fc(\tau',x,p)\,dp\,\,dS_r\,d\tau'.
\]
As in the proof of (3) of Proposition \ref{evolutionN}, the uniform estimate (\ref{estP}) implies that the integral in the right hand side vanishes for $r$ large enough, whence the identity $M_2(\tau)=M_1(\tau)$ follows by letting $r\to\infty$ in the previous equation.\prfe
\begin{Remark}\textnormal{
From a physics point of view, it would be desirable that the emission of outgoing energy is a {\it generic} property of isolated solutions. We conjecture that spherically symmetric solutions are the only isolated solutions of the Vlasov-Maxwell system for which the energy $M_2(\tau)$ is constant.}
\end{Remark}

\section{The one dimensional system}\label{1d}
\setcounter{equation}{0} 
In this section we consider the Vlasov-Maxwell system in one spatial dimension and two dimensions in velocity, which is also known as the ``one and one-half dimensional Vlasov-Maxwell system". 
We remark that, besides greatly reducing the mathematical complexity of the problem, one dimensional Vlasov-Maxwell systems have applications in plasma physics---e.g., to model laser-plasma interactions or electron beams---and in numerical simulations, see \cite{FGS} and the references therein.
In Cartesian coordinates, the Vlasov-Maxwell system in one spatial dimension and two dimensions in velocity takes the form
\begin{equation}\label{vlasov}
\partial_t f+\widehat{p}_1\partial_x f+(E_1+B\widehat{p}_2)\partial_{p_1}f+(E_2-B\widehat{p}_1)\partial_{p_2}f=0,
\end{equation}
\begin{equation}
\partial_t E_1=-j_1,\quad \partial_x E_1=\rho-n,\label{maxwell11d}
\end{equation}
\begin{equation}
\partial_t E_2+\partial_xB=-j_2,\quad \partial_tB+\partial_xE_2=0,\label{maxwell21d}
\end{equation}
\begin{equation}\label{sources}
(\rho,j_1,j_2)=\int (1,\widehat{p_1},\widehat{p_2})f\,dp,
\end{equation}
see \cite{GSh2}.
The particle density in phase-space is the function $f=f(t,x,p)$, where $(t,x)\in\R^2$ denotes a Cartesian system of coordinates in the two-dimensional Minkowski space, $p=(p_1,p_2)\in\R^2$ is the particles momentum and, for all $z\in\R$, we set
\[
\widehat{z}=\frac{z}{\sqrt{1+z^2}}.
\]
In particular, $\widehat{p}=(\widehat{p}_1,\widehat{p}_2)$ is the relativistic velocity. 
The electric field is given by $E=(E_1,E_2)=E(t,x)$, while the magnetic field is the scalar function $B=B(t,x)$. 
The function $n=n(x)$ is a background density that will be chosen in such a way that the component $E_1$ of the electric field vanishes at infinity (the latter condition is necessary for the existence of solutions with finite energy). 
The continuity equation and the energy identity for the system (\ref{vlasov})--(\ref{sources}) take the form
\[
\partial_t\rho+\partial_xj_1=0, \quad \partial_t e+\partial_x\mathfrak{p}=0
\] 
where
\[
e(t,x)=\frac{1}{2}|E|^2+\frac{1}{2}B^2+\int\sqrt{1+|p|^2}f\,dp,\quad
\mathfrak{p}(t,x)=E_2B+\int p_1f\,dp.
\]
As in the three dimensional case, we start by reformulating the system (\ref{vlasov})--(\ref{maxwell21d}) in terms of new dynamical variables evaluated on the backward hyperboloids $t+\sqrt{1+x^2}=\tau$, so that the limit at past null infinity becomes $x\to\pm\infty$; define
\[
\fc(\tau,x,p)= f(\tau-\sqrt{1+x^2},x,p), \quad (\rhoc,\jc)(\tau,x)=(\rho,j)(\tau-\sqrt{1+x^2},x),
\]
\[
\Uc(\tau,x)=E_1(\tau-\sqrt{1+x^2},x),\quad (\phic,\psic)=\frac{1}{2}\left(E_2+B,E_2-B\right)(\tau-\sqrt{1+x^2},x).
\]
The system (\ref{vlasov})--(\ref{maxwell21d}) in terms of these new unknowns reads
\begin{eqnarray}
&&(1+\widehat{p}_1\widehat{x})\partial_\tau\fc+\widehat{p}_1\partial_x\fc+[\Uc+(\phic-\psic)\widehat{p}_2]\partial_{p_1}\fc\nonumber\\
&&\quad\quad+[\phic(1-\widehat{p}_1)+\psic(1+\widehat{p}_1)]\partial_{p_2}\fc=0,\label{vlasova}
\end{eqnarray}
\begin{equation}\label{maxwell1a}
\partial_\tau \Uc=-\jc^1,\quad\partial_x\Uc=\rhoc+\jc^1\widehat{x}-n,
\end{equation}
\begin{equation}\label{maxwell2a}
\left(\partial_\tau+\frac{1}{1+\widehat{x}}\partial_x\right)\phi=-\frac{\jc^2}{2(1+\widehat{x})},\quad
\left(\partial_\tau-\frac{1}{1-\widehat{x}}\partial_x\right)\psi=-\frac{\jc^2}{2(1-\widehat{x})}.
\end{equation}
The NIRC in the future for the system (\ref{vlasova})--(\ref{maxwell2a}) is
\begin{equation}\label{NIRC1d}
\lim_{r\to\infty}\int_{\tau_1}^{\tau_2}\left[\left(\phic^2-\psic^2\right)(\tau,-r)-\left(\phic^2-\psic^2\right)(\tau,r)\right]\,dx=0\quad\forall\tau_1,\tau_2\geqslant 0.
\end{equation}
The main result of this section is a global (in the future) existence and uniqueness theorem of classical solutions to the system (\ref{vlasova})--(\ref{maxwell2a}). Moreover a necessary and sufficient condition on the data of the problem is given such that the solutions satisfy (\ref{NIRC1d}). In order to give a precise formulation of our results, we need first to introduce some preliminaries and notation. Firstly we shall assume throughout that the background density $n(x)$ is smooth with compact support and is neutralizing, namely
\begin{equation}\label{neutral}
\int_{\R}\left[(\rho+\jc^1\widehat{x})(0,x)-n(x)\right]\,dx=0.
\end{equation}
This is the only possibility which leads to a finite energy solution, see Remark \ref{remark} .
Let $(\ec,\pc)(\tau,x)=(e,\mathfrak{p})(\tau-\sqrt{1+x^2},x)$; the energy identity takes the form
\begin{equation}\label{enident1d}
\partial_\tau(\ec+\pc\widehat{x})=-\partial_x\pc.
\end{equation}
The mass-energy on the backward hyperboloids in one spatial dimension are given by
\[
N_0(\tau)=\lim_{r\to\infty}n_0(\tau,r),\quad n_0(\tau,r)=\int_{-r}^r\left(\rhoc+\jc^1\widehat{x}\right)(\tau,x)\,dx,
\]
\[
M_0(\tau)=\lim_{r\to\infty}m_0(\tau,r),\quad m_0(\tau,r)=\int_{-r}^r\left(\ec+\pc\widehat{x}\right)(\tau,x)\,dx,
\]
and as before we set $N_0(0)=N_0^{\mathrm{in}}$ and $M_0(0)=M_0^{\mathrm{in}}$. Note that
\[
M_0^{\mathrm{in}}=\int\int p_0\fcin\,dx\,dp+\int\left[\frac{1}{2}\Uc^2(0,x)+\left(\phicin\right)^2(1+\widehat{x})+\left(\psicin\right)^2(1-\widehat{x})\right]\,dx,
\]
where we denoted 
\[
p_0=\sqrt{1+|p|^2}+p_1\widehat{x},\quad \left(\fcin,\phicin,\psicin\right)=\left(\fc,\phic,\psic\right)_{|\tau=0}.
\]
For $T>0$, let $\mathcal{C}_T$ denote the set of quadruples $(\fc,\Uc,\phic,\psic)$ of functions
\[
\fc:[0,T)\times\R\times\R^2\to [0,\infty),\quad \Uc,\phic,\psic:[0,T)\times\R\to\R,
\]
with regularity $\fc,\Uc,\phic,\psic\in C^1$ and such that
\begin{equation}\label{momentumsupport}
\mathcal{P}_\cap(\tau)=\sup\{\sqrt{1+|p|^2}:f(s,x,p)\neq 0,\,s\in [0,\tau],\,x\in\R\}<\infty,\quad\forall\tau\in [0,T).
\end{equation}
The solutions of (\ref{vlasova})--(\ref{maxwell2a}) that we construct belong to the class $\mathcal{C}_\infty$. This set of solutions is the analogue of what we used in the three dimensional case. In particular, for solutions in the class $\mathcal{C}_\infty$ the results of Section \ref{generalproperties} hold with the obvious modifications in the one dimensional case. 
In order to obtain a unique solution in $\mathcal{C}_\infty$, an initial data set has to be prescribed. We assume
\begin{itemize}
\item[(I)] $\fcin:\R\times\R^2\to[0,\infty)$ such that $\fcin\in C^1_c$; 
\item[(II)] $\phicin,\psicin:\R\to\R$ such that $\phicin,\psicin\in C^1_b$;  
\item[(III)] $\lim_{x\to+\infty}\phicin=\lim_{x\to-\infty}\psicin=0$;
\item[(IV)] The limits 
\[
\lim_{x\to -\infty}\phicin,\quad\lim_{x\to +\infty}\psicin,\quad \lim_{x\to -\infty}\frac{{\phicin}'}{1+\widehat{x}},\quad\lim_{x\to +\infty}\frac{{\psicin}'}{1-\widehat{x}}
\]
exist and are bounded.
\end{itemize}
%The notation used above has the following meaning: The index $b$ means that all derivatives up to the indicated order are
%bounded and a prime idenotes the differentiation of functions of one real variable. 
\begin{Remark}\label{remark}
\textnormal{Once the limits in (III) are required to exist, their vanishing is necessary for the initial data to satisfy $M_0^{\mathrm{in}}<\infty$. For the same reason we must have $\lim_{x\to\pm\infty}U(0,x)=0$ and so, by the second equation in (\ref{maxwell1a}), 
\[
\Uc(0,x)=\int_{-\infty}^x\left [(\rhoc+\jc^1\widehat{x})(0,x')-n(x')\right]\,dx',
\]
whence (\ref{neutral}) is necessary for having $\lim_{x\to +\infty}U(0,x)=0$. Note also that $\Uc(0,x)$ is compactly supported.}
\end{Remark}

Besides initial conditions, uniqueness of solutions also requires data at past null infinity, which we take as 
\begin{itemize}
\item[(V)] $\phic^-,\psic^+:[0,\infty)\to\R$  such that $\phic^-,\psic^+\in C^1_b\cap L^2$.
\end{itemize}
\begin{Definition} 
A data set $\{\fcin,\phicin,\psicin,\phic^-,\psic^+\}$ is said to be \textnormal{admissible} if it satisfies (I)--(V) and the conditions 
\begin{eqnarray*}
&&\textnormal{(VI)}\quad \lim_{x\to -\infty}\phicin(x)=\phic^-(0),\quad \lim_{x\to +\infty}\psicin(x)=\psic^+(0),\\
&&\textnormal{(VII)}\quad \lim_{x\to -\infty}\frac{{\phicin}'(x)}{1+\widehat{x}}=-{\phic^-}'(0),\quad
\lim_{x\to +\infty}\frac{{\psicin}'(x)}{1-\widehat{x}}={\psic^-}'(0).
\end{eqnarray*}
\end{Definition}
Our main result is the following theorem.
\begin{Theorem}\label{main1}
Let an admissible data set $\{\fcin,\phicin,\psicin,\phic^-,\psic^+\}$ be given such that $M_0^{\mathrm{in}}<\infty$. Then there exists a unique $(\fc,\Uc,\phic,\psic)\in\mathcal{C}_\infty$ global solution of (\ref{vlasova})--(\ref{maxwell2a}) such that
\begin{itemize}
\item[(i)] $\fc(0,x,p)=\fcin(x,p),\quad (\phic,\psic)(0,x)=(\phicin,\psicin)(x),\quad\forall x\in\R,\,p\in\R^2$;
\item[(ii)] $\lim_{x\to -\infty}\phic(\tau,x)=\phic^-(\tau),\ \lim_{x\to +\infty}\psic(\tau,x)=\psic^+(\tau),\quad\forall\tau\geqslant 0$.
\end{itemize}
\end{Theorem}

The solution of Theorem 1 is not always isolated. This depends on the choice of the data, according to the following theorem.
\begin{Theorem}\label{main2}
For the solution of Theorem \ref{main1} we have
\begin{equation}\label{M01d}
M_0(\tau)=M_0^{\mathrm{in}}+\int_0^\tau\left[\left(\phic^-\right)^2+\left(\psic^+\right)^2\right](\tau')\,d\tau'.
\end{equation}
Moreover the solution satisfies (\ref{NIRC1d}), and thus it is isolated, if and only if $\phic^-(\tau)=\psi^+(\tau)=0$, for all $\tau\geqslant 0$. 
\end{Theorem}

We split the proof of the above results in five steps. In Section \ref{representationformula} we derive an integral representation formula for the fields by integrating along the characteristics of the operators $\partial_\tau\pm(1\pm\widehat{x})^{-1}\partial_x$ in the $(\tau,x)$-plane. The proof of Theorem \ref{main2} relies only upon this formula and is presented in Section \ref{proof2}. In Section \ref{estfield} we estimate local solutions in $L^\infty$. Unlike the case of the Cauchy problem considered in \cite{GSh2}, in the present situation we have to consider separately four different regions of the plane. In each of these regions, a uniform bound on the fields is obtained by integrating the energy identity over suitable subsets of the $(\tau,x)$-plane. In Section \ref{local} we formulate a local existence and uniqueness theorem together with a continuation criterium which states the a local solution can blow-up in finite time only if the momentum support of the particle density becomes unbounded. The proof of this result can be obtained by standard methods and is therefore omitted. In Section \ref{boundmomentum} the momentum support of the particle density is estimated, thereby completing the proof of Theorem \ref{main1}.   

\subsection{A representation formula for solutions in the class $\mathcal{C}_\infty$}\label{representationformula}
Recall that $p_0=\sqrt{1+|p|^2}+p_1\widehat{x}$.
Since $p_0>0$ in the support of $\fc$ under the assumption (\ref{momentumsupport}), the Vlasov equation for solutions in the class $\mathcal{C}_\infty$ can be rewritten as
\begin{eqnarray}\label{vlasovb}
&&\partial_\tau\fc+\frac{{p}_1}{p_0}\partial_x\fc+\frac{1}{p_0}[\sqrt{1+|p|^2}\Uc+(\phic-\psic)p_2]\partial_{p_1}f\nonumber\\
&&\quad\quad+\frac{1}{p_0}[\phic(\sqrt{1+|p|^2}-p_1)+\psic(\sqrt{1+|p|^2}+p_1)]\partial_{p_2}f=0.
\end{eqnarray}
The solution of (\ref{vlasovb}) with initial datum $\fcin$ is given by $\fc(\tau,x,p)=\fcin(X(0),P(0))$,
where $(X(s),P(s))$ is the solution of the characteristic system
\begin{equation}\label{char1da}
\dot{x}=\frac{{p}_1}{p_0},\quad\dot{p}_1=\frac{1}{p_0}[\sqrt{1+|p|^2}\Uc+(\phic-\psic)p_2],
\end{equation}
\begin{equation}\label{char1db}
\dot{p}_2=\frac{1}{p_0}[\phic(\sqrt{1+|p|^2}-p_1)+\psic(\sqrt{1+|p|^2}+p_1)]
\end{equation}
subject to the condition $(X(t),P(t))=(x,p)$. It follows by Lemma \ref{supportf} that the subset of the $(\tau,x)$-plane defined by 
\[
\mathcal{V}=\{(\tau,x):\tau\geqslant 0,\,\sqrt{1+x^2}\geqslant R_0+\frac{1}{2}\tau\},
\]
is a vacuum region, i.e., $\fc=0$, for $(\tau,x,p)\in\mathcal{V}\times\R^2$.

Next we derive an integral representation formula for solutions of (\ref{maxwell2a}) having regularity, initial data and decay as stated in Theorem \ref{main1}. 
To this purpose it is convenient to introduce the auxiliary functions
\[
g_+(z)=\frac{z^2-1}{2z},\quad g_-(z)=-g_+(z),\quad z>0.
\] 
Note that
\[
g'_{\pm}(z)=\pm\frac{1+z^2}{2z^2}=\frac{\pm 1}{1\pm\widehat{g}_\pm(z)},\quad z>0.
\]
Whence $g_{\pm}$ are invertible functions and we have
\[
g^{-1}_{\pm}(z)=\sqrt{1+z^2}\pm z,\quad z\in\R.
\]
Consider the following subsets of the half-plane $\tau\geqslant 0$:
\[
\Omega_1=\{(\tau,x):0\leqslant\tau<\min\left(g^{-1}_-(x),g^{-1}_+(x)\right)\},\quad 
\Omega_2=\{(\tau,x):g^{-1}_-(x)\leqslant\tau<g^{-1}_+(x)\},
\]
\[
\Omega_3=\{(\tau,x):\tau\geqslant\max\left(g^{-1}_-(x),g^{-1}_+(x)\right)\},\quad\Omega_4=\{(\tau,x):g^{-1}_+(x)\leqslant\tau<g^{-1}_-(x)\}.
\]
\begin{Proposition}\label{reprfields}
Let $\jc^2\in C^1$ be given such that $\jc^2=0$, for $(\tau,x)\in\mathcal{V}$ and let $\phicin,\psicin,\phic^-,\psic^+$ satisfying (II)--(VII) be given. There exists a unique $C^1$ solution $(\phic,\psic)$ of (\ref{maxwell2a}) satisfying 
\begin{itemize}
\item[(A)] $(\phic,\psic)_{|t=0}=(\phicin,\psicin)$,
\item[(B)] $\lim_{x\to -\infty}\phic(\tau,x)=\phic^-(\tau)$, $\lim_{x\to +\infty}\psic(\tau,x)=\psic^+(\tau)$, for all $\tau\geqslant 0$.
\end{itemize}
Moreover the following holds:
\begin{itemize}
\item[(i)] $\phic(\tau,x)$ is given by
\begin{eqnarray}
\phic(\tau,x)&=&\left(\phicin\circ g_+\right)\left(g^{-1}_+(x)-\tau\right)\nonumber\\
&&-\frac{1}{2}\int_{g_+\left(g^{-1}_+(x)-\tau\right)}^x\jc^2\left(g^{-1}_+(y)-g^{-1}_+(x)+\tau,y\right)\,dy,\label{reprphi1}
\end{eqnarray}
for $(\tau,x)\in\Omega_1\cup\Omega_2$ and by
\begin{eqnarray}
\phic(\tau,x)&=&\phic^-\left(\tau-g^{-1}_+(x)\right)\nonumber\\
&&-\frac{1}{2}\int_{-\infty}^x\jc^2\left(g^{-1}_+(y)-g^{-1}_+(x)+\tau,y\right)\,dy,\label{reprphi2}
\end{eqnarray}
for $(\tau,x)\in\Omega_3\cup\Omega_4$.
\item[(ii)] $\psic(\tau,x)$ is given by
\begin{eqnarray}
\psic(\tau,x)&=&\left(\psicin\circ g_-\right)\left(g^{-1}_-(x)-\tau\right)\nonumber\\
&&-\frac{1}{2}\int^{g_-\left(g^{-1}_-(x)-\tau\right)}_x\jc^2\left(g^{-1}_-(y)-g^{-1}_-(x)+\tau,y\right)\,dy,\label{reprpsi1}
\end{eqnarray}
for $(\tau,x)\in\Omega_1\cup\Omega_4$ and by
\begin{eqnarray}
\psic(\tau,x)&=&\psic^+\left(\tau-g^{-1}_-(x)\right)\nonumber\\
&&-\frac{1}{2}\int^{+\infty}_x\jc^2\left(g^{-1}_-(y)-g^{-1}_-(x)+\tau,y\right)\,dy,\label{reprpsi2}
\end{eqnarray}
for $\Omega_2\cup\Omega_3$.
\item[(iii)] In the vacuum region $\mathcal{V}$ we have
\[
\phic(\tau,x)=\left\{\begin{array}{ll}\left(\phicin\circ g_+\right)\left(g^{-1}_+(x)-\tau\right),&\textnormal{ {\it for} }(\tau,x)\in\left(\Omega_1\cup\Omega_2\right)\cap\mathcal{V},\\
\phic^-\left(\tau-g^{-1}_+(x)\right)&\textnormal{ {\it for} }(\tau,x)\in\left(\Omega_3\cup\Omega_4\right)\cap\mathcal{V},
\end{array}\right.
\]
\[
\psic(\tau,x)=\left\{\begin{array}{ll}\left(\psicin\circ g_-\right)\left(g^{-1}_-(x)-\tau\right),&\textnormal{ {\it for} }(\tau,x)\in\left(\Omega_1\cup\Omega_4\right)\cap\mathcal{V},\\
\psic^+\left(\tau-g^{-1}_-(x)\right)&\textnormal{ {\it for} }(\tau,x)\in\left(\Omega_2\cup\Omega_3\right)\cap\mathcal{V},
\end{array}\right.
\]
\item[(iv)] The following limits
\[
\lim_{x\to +\infty}\phic(\tau,x)=0,\quad\lim_{x\to -\infty}\psic(\tau,x)=0,
\]
\[
\lim_{x\to -\infty}\phic(\tau,x)=\phic^-(\tau),\quad\lim_{x\to+\infty}\psic(\tau,x)=\psic^+(\tau),
\]
are attained uniformly in compact subsets of $\tau\geqslant 0$.
\end{itemize}
\end{Proposition}
\noindent\textit{Proof: }We give the details of the proof for $\phic$ only, the argument for $\psic$ being the same. Let us show first that the function $\phic$ defined by (\ref{reprphi1})-(\ref{reprphi2}) is $C^1$. Owing to $\jc^2(\tau,x)=0$, for $\sqrt{1+x^2}\geqslant R_0+\frac{1}{2}\tau$, it is easy to see that the integrals in the right hand sides of (\ref{reprphi1})-(\ref{reprphi2}) are extended over compact sets and so $\phic$ is $C^1$ in the interior of each of the two regions $\Omega_1\cup\Omega_2$ and $\Omega_3\cup\Omega_4$. The border between these regions is the curve $\tau=g_+^{-1}(x)$ and we have
\[
\lim_{\tau\to g_+^{-1}(x)^-}\phic(\tau,x)=\lim_{x\to -\infty}\phicin(x)-\frac{1}{2}\int_{-\infty}^x\jc^2(g^{-1}(y),y)\,dy,
\]
\[
\lim_{\tau\to g_+^{-1}(x)^+}\phic(\tau,x)=\phic^-(0)-\frac{1}{2}\int_{-\infty}^x\jc^2(g^{-1}(y),y)\,dy.
\]
Hence the continuity of $\phic$ follows by (VI).
We emphasize that the integral in the right hand side of the last two equations is finite, as $\jc^2(g_+^{-1}(y),y)=0$, for $y\leqslant g_-(2R_0)$.
Taking a $\tau$-derivative in (\ref{reprphi1}) and (\ref{reprphi2}) we obtain
\begin{eqnarray*}
\partial_\tau\phic&=&-\left(\frac{{\phicin}'\circ g_+}{1+\widehat{g}_+(\cdot)}\right)\left(g_+^{-1}(x)-\tau\right)-\frac{1}{2}g'_+\left(g_+^{-1}(x)-\tau\right)\jc^2\left(0,g_+\left(g_+^{-1}(x)-\tau\right)\right)\nonumber\\
&&-\frac{1}{2}\int_{g_+\left(g^{-1}_+(x)-\tau\right)}^x\partial_\tau\jc^2\left(g^{-1}_+(y)-g^{-1}_+(x)+\tau,y\right)\,dy,
\end{eqnarray*}
for $(\tau,x)\in\Omega_1\cup\Omega_2$ and
\[
\partial_\tau\phic={\phic^-}'\left(\tau-g_+^{-1}(x)\right)-\frac{1}{2}\int_{-\infty}^x\partial_\tau\jc^2\left(g^{-1}_+(y)-g^{-1}_+(x)+\tau,y\right)\,dy,
\]
for $(\tau,x)\in\Omega_3\cup\Omega_4$.  Using again the support property of $\jc^2$ and (VII), we find that $\lim_{\tau\to g_+^{-1}(x)^-}\partial_\tau\phic=\lim_{\tau\to g_+^{-1}(x)^+}\partial_\tau\phic$.
In the same way one can show that $\partial_x\phic$ is continuous. A straightforward computation reveals that $\phic$ solves the first equation in (\ref{maxwell2a}) with the initial-decay conditions $\phic(0,x)=\phicin(x)$, $\lim_{x\to -\infty}\phic(\tau,x)=\phic^-(\tau)$. Let us show now that any solution to the latter problem has the form (\ref{reprphi1})-(\ref{reprphi2}), thereby proving the uniqueness part of the theorem. Let $\phic$ be a $C^1$ solution of the first equation in (\ref{maxwell2a}) with the initial-decay conditions $\phic(0,x)=\phicin(x)$, $\lim_{x\to -\infty}\phic(\tau,x)=\phic^-(\tau)$. It follows that the function $h(\tau,z)=\phic(\tau,g_+(z)),\,z>0$, satisfies the problem 
\begin{equation}\label{temp1}
(\partial_\tau+\partial_z)h=-\frac{g_+'}{2}\mu, \quad h(0,z)=\left(\phicin\circ g_+\right)(z),\quad \lim_{z\to 0^+}h(\tau,z)=\phic^-(\tau),
\end{equation}
where $\mu(\tau,z)=\jc^2(\tau,g_+(z))$. Integrating along characteristics one finds that the solution of (\ref{temp1}) is 
\[
h(\tau,z)= \left(\phicin\circ g_+\right)(z-\tau)-\frac{1}{2}\int_0^\tau \left(\mu(s,\cdot)g'_+\right)(z+s-\tau)\,ds, 
\]
for $\tau<z$ and
\[
h(\tau,z)= \phic^-(\tau-z)-\frac{1}{2}\int_{\tau-z}^\tau\left(\mu(s,\cdot)g'_+\right)(z+s-\tau)\,ds, 
\]
for $\tau\geqslant z$. As $\mu(\tau,z)=\jc^2(\tau,g_+(z))$ and by a change of variable in the integrals we obtain
\[
h(\tau,z)= \left(\phicin\circ g_+\right)(z-\tau)-\frac{1}{2}\int_{g_+(z-\tau)}^{g_+(z)}\jc^2(g_+^{-1}(y)+\tau-z,y)\,dy, 
\]
for $\tau<z$ and
\[
h(\tau,z)= \phic^-(\tau-z)-\frac{1}{2}\int_{-\infty}^{g_+(z)}\jc^2(g_+^{-1}(y)+\tau-z,y)\,dy, 
\]
for $\tau\geqslant z$. Since $\phic(\tau,x)=h(\tau,g_+^{-1}(x))$, this completes the proof of the representation formula for $\phic$. 
Next we prove (iii). Since $\jc^2(\tau,x)=0$, for $\sqrt{1+x^2}\geqslant R_0+\frac{1}{2}\tau$, we have
\begin{equation}\label{temp}
\jc^2\left(g^{-1}_+(y)-g^{-1}_+(x)+\tau,y\right)=0
\end{equation}
for $\frac{1}{2}g^{-1}_-(y)\geqslant R_0+\frac{1}{2}\tau-\frac{1}{2}g_+^{-1}(x)$. Hence for $y\leqslant x$, (\ref{temp}) is satisfied if
\[
\frac{1}{2}g_-^{-1}(x)\geqslant R_0+\frac{1}{2}\tau-\frac{1}{2}g^{-1}_+(x),
\]
i.e., if $\sqrt{1+x^2}=\frac{1}{2}(g_-^{-1}(x)+g_+^{-1}(x))\geqslant R_0+\frac{1}{2}\tau$. Using this in (\ref{reprphi1})-(\ref{reprphi2}) concludes the proof of the claim for $\phic$. By (iii) and the condition (III) on the initial data, the proof of (iv) is straightforward.\prfe
\begin{Remark}\textnormal{When $\phic^-=\psic^+=0$, i.e., when---according to Theorem \ref{main2}---the solution is isolated, the formulas (\ref{reprphi2}), (\ref{reprpsi2}) imply that the field in the region $\Omega_3$ is given by the retarded solution of the equations.}
\end{Remark}
\subsection{Proof of Theorem \ref{main2}}\label{proof2}
In this subsection we prove Theorem \ref{main2}. By (iv) of Proposition \ref{reprfields}, the left hand side of (\ref{NIRC1d}) equals 
\[
\int_{\tau_1}^{\tau_2}\left[\left(\phic^-\right)^2+\left(\psic^+\right)^2\right]\,d\tau,
\]
which vanishes for all $\tau_1,\tau_2\geqslant 0$ if and only if $\phic^-\equiv \psic^+\equiv 0$. The analogue of (\ref{m0}) in one spatial dimension is
\[
m_0(\tau,r)=m_0(0,r)+\int_0^\tau\left[\pc(\tau',-r)-\pc(\tau',r)\right]\,d\tau'.
\]
For $r$ sufficientely large, the integral in the right hand side equals
\[
\int_{0}^{\tau}\left[\left(\phic^2-\psic^2\right)(\tau,-r)-\left(\phic^2-\psic^2\right)(\tau,r)\right]\,dx.
\]
Letting $r\to +\infty$ and using again (iv) of Proposition \ref{reprfields} concludes the proof of (\ref{M01d}).

\subsection{Bounds on the fields}\label{estfield}
In this subsection we prove that the fields are uniformly bounded. Let $(\fc,\Uc,\phic,\psic)\in\mathcal{C}_T$ be a local solution of (\ref{vlasova})--(\ref{maxwell2a}) which match the initial-decay data. Then
\begin{eqnarray*}
|\Uc(\tau,x)|&=&\left|\int_{-\infty}^x\left(\rhoc+\jc^1\widehat{x}\right)(\tau,x')\,dx'-\int_{-\infty}^xn(x')\,dx'\right|\\
&\leqslant& N_0(\tau)+\|n\|_{L^1}=N_0^{\mathrm{in}}+C\leqslant C\quad \forall\, \tau\in [0,T),
\end{eqnarray*}
whence $\|\Uc(\tau)\|_\infty\leqslant C$, for all $\tau\in [0,T)$.
Here and in the following we denote by $C$ a generic positive constant which may change from line to line. The estimation of the fields $\phic,\psic$ is not so simple. We shall need the following
\begin{Lemma}\label{tool}
\[
\left(\ec\pm\pc\right)(\tau,x)\geqslant |\jc^2|(\tau,x),\quad\forall\,(\tau,x)\in [0,\infty)\times\R.
\]
\end{Lemma}
\noindent\textit{Proof: }From one hand
\[
|\jc^2|\leqslant \int\frac{|p_2|}{\sqrt{1+|p|^2}}\fc\,dp;
\]
on the other hand
\begin{eqnarray*}
\ec\pm\pc&=&\int\left(\sqrt{1+|p|^2}\pm p_1\right)\fc\,dp+\frac{1}{2}\Uc^2+\phic^2+\psic^2\pm\phic^2\mp\psic^2\\
&\geqslant&\int\left(\sqrt{1+|p|^2}\pm p_1\right)\fc\,dp.
\end{eqnarray*}
The lower bound
\[
\sqrt{1+|p|^2}\pm p_1=\frac{1+|p|^2-p_1^2}{\sqrt{1+|p|^2}\mp p_1}\geqslant\frac{1+p_2^2}{2\sqrt{1+|p|^2}}\geqslant\frac{|p_2|}{\sqrt{1+|p|^2}},
\]
yields the result.\prfe
\begin{Proposition}
Let $(\fc,\Uc,\phic,\psic)\in\mathcal{C}_T$ be a local solution of (\ref{vlasova})--(\ref{maxwell2a}) with admissible initial-decay data such that $M_0^{\mathrm{in}}<\infty$. Then 
\[
\|\phic(\tau)\|_\infty+\|\psic(\tau)\|_\infty\leqslant C,\quad\forall\, \tau\in [0,T).
\]
\end{Proposition}
\noindent\textit{Proof: }By Lemma \ref{tool} and (\ref{reprphi1})--(\ref{reprpsi2}) we have 
\begin{equation}\label{estphi1}
|\phic(\tau,x)|\leqslant C+\frac{1}{2}\int_{g_+\left(g^{-1}_+(x)-\tau\right)}^x\left(\ec-\pc\right)\left(g^{-1}_+(y)-g^{-1}_+(x)+\tau,y\right)\,dy,
\end{equation}
for $(\tau,x)\in\Omega_1\cup\Omega_2$ and 
\begin{equation}\label{estphi2}
|\phic(\tau,x)|\leqslant C+\frac{1}{2}\int_{-\infty}^x\left(\ec-\pc\right)\left(g^{-1}_+(y)-g^{-1}_+(x)+\tau,y\right)\,dy,
\end{equation}
for $(\tau,x)\in\Omega_3\cup\Omega_4$. At the same fashion,
\begin{equation}\label{estpsi1}
|\psic(\tau,x)|\leqslant C+\frac{1}{2}\int^{g_-\left(g^{-1}_-(x)-\tau\right)}_x\left(\ec+\pc\right)\left(g^{-1}_-(y)-g^{-1}_-(x)+\tau,y\right)\,dy,
\end{equation}
for $(\tau,x)\in \Omega_1\cup\Omega_4$ and 
\begin{equation}\label{estpsi2}
|\psic(\tau,x)|\leqslant C+\frac{1}{2}\int^{+\infty}_x\left(\ec+\pc\right)\left(g^{-1}_-(y)-g^{-1}_-(x)+\tau,y\right)\,dy,
\end{equation}
for $(\tau,x)\in \Omega_2\cup\Omega_3$. Thus it is enough to estimate the integrals in the right hand sides of (\ref{estphi1})--(\ref{estpsi2}). For $(\tau_0,x_0)\in\Omega_1$, we integrate the energy identity (\ref{enident1d}) in the interior of the past light cone with vertex on $(\tau_0,x_0)$, namely the region of the $(\tau,x)$-plane so defined:
\[
\Delta(\tau_0,x_0)=\left\{(\tau,x):0\leqslant \tau\leqslant \min\left(g_+^{-1}(x)+\tau_0-g_+^{-1}(x_0),g_-^{-1}(x)+\tau_0-g_-^{-1}(x_0)\right)\right\}.
\]
So doing we obtain 
\begin{eqnarray}
&&\int_{g_+\left(g^{-1}_+(x_0)-\tau_0\right)}^{x_0}\left(\ec-\pc\right)\left(g^{-1}_+(x)-g^{-1}_+(x_0)+\tau_0,x\right)\,dx\nonumber\\
&&\quad\quad+\int^{g_-\left(g^{-1}_-(x_0)-\tau_0\right)}_{x_0}\left(\ec+\pc\right)\left(g^{-1}_-(x)-g^{-1}_-(x_0)+\tau_0,x\right)\,dx\nonumber\\
&&\quad=\int_{g_+\left(g_+^{-1}(x_0)-\tau_0\right)}^{g_-\left(g_-^{-1}(x_0)-\tau_0\right)}\left(\ec+\pc\widehat{x}\right)(0,x)\,dx\leqslant M_0^{\mathrm{in}}\label{uela1},\quad\forall\,(\tau_0,x_0)\in\Omega_1.
\end{eqnarray}
For $(\tau_0,x_0)\in\Omega_2$, we integrate (\ref{enident1d}) in the interior of the truncated cone
\[
\Delta_{(R)}(\tau_0,x_0)=\Delta(\tau_0,x_0)\cap\left\{(\tau,x):g_+\left(g_+^{-1}(x_0)-\tau_0\right)\leqslant x\leqslant R\right\},\quad R>x_0.
\]
So doing we get
\begin{eqnarray}
&&\int_{g_+\left(g^{-1}_+(x_0)-\tau_0\right)}^{x_0}\left(\ec-\pc\right)\left(g^{-1}_+(x)-g^{-1}_+(x_0)+\tau_0,x\right)\,dx\nonumber\\
&&\quad\quad+\int^{R}_{x_0}\left(\ec+\pc\right)\left(g^{-1}_-(x)-g^{-1}_-(x_0)+\tau_0,x\right)\,dx\label{uela2}\\
&&\quad=\int_{g_+\left(g_+^{-1}(x_0)-\tau_0\right)}^{R}\left(\ec+\pc\widehat{x}\right)(0,x)\,dx-\int_0^{g_-^{-1}(R)+\tau_0-g_-^{-1}(x_0)}\pc(\tau,R)\,d\tau.\nonumber
\end{eqnarray}
By (iv) of Proposition \ref{reprfields} we have
\[
-\int_0^{g_-^{-1}(R)+\tau_0-g_-^{-1}(x_0)}\pc(\tau,R)\,d\tau\longrightarrow\int_0^{\tau_0-g_-^{-1}(x_0)}\left(\psic^+\right)^2(\tau)\,d\tau, \ \textnormal{ as } R\to\infty.
\]
Whence in the limit $R\to\infty$, (\ref{uela2}) entails
\begin{eqnarray}
&&\int_{g_+\left(g^{-1}_+(x_0)-\tau_0\right)}^{x_0}\left(\ec-\pc\right)\left(g^{-1}_+(x)-g^{-1}_+(x_0)+\tau_0,x\right)\,dx\nonumber\\
&&\quad\quad +\int^{\infty}_{x_0}\left(\ec+\pc\right)\left(g^{-1}_-(x)-g^{-1}_-(x_0)+\tau_0,x\right)\,dx\nonumber\\
&&\quad\leqslant M_0^{\mathrm{in}}+\|\psic^+\|_{L^2},\quad\forall\,(\tau_0,x_0)\in\Omega_2.\label{uela3}
\end{eqnarray}
A similar argument can be applied in the regions $\Omega_3$ and $\Omega_4$. Precisely, for $(\tau_0,x_0)\in\Omega_3$ the energy identity (\ref{enident1d}) is integrated in the truncated cone
\[
\Delta_{(-R,R)}(\tau_0,x_0)=\Delta(\tau_0,x_0)\cap\left\{(\tau,x):  |x-x_0|\leqslant R\right\},\quad R>0,
\]
whereas for $(\tau_0,x_0)\in\Omega_4$ we choose 
\[
\Delta_{(-R)}(\tau_0,x_0)=\Delta(\tau_0,x_0)\cap\left\{(\tau,x):-R\leqslant x\leqslant g_-\left(g_-^{-1}(x_0)-\tau_0\right)\right\},\quad R>-x_0.
\]
So doing and letting $R\to\infty$ we arrive to the estimates 
\begin{eqnarray}
&&\int_{-\infty}^{x_0}\left(\ec-\pc\right)\left(g^{-1}_+(x)-g^{-1}_+(x_0)+\tau_0,x\right)\,dx\nonumber\\
&&\quad\quad+\int^{+\infty}_{x_0}\left(\ec+\pc\right)\left(g^{-1}_-(x)-g^{-1}_-(x_0)+\tau_0,x\right)\,dx\nonumber\\
&&\quad\leqslant M_0^{\mathrm{in}}+\|\phic^-\|_{L^2}+\|\psic^+\|_{L^2},\quad \forall\,(\tau_0,x_0)\in\Omega_3\label{uela4},
\end{eqnarray}
\begin{eqnarray}
&&\int_{-\infty}^{x_0}\left(\ec-\pc\right)\left(g^{-1}_+(x)-g^{-1}_+(x_0)+\tau_0,x\right)\,dx\nonumber\\
&&\quad\quad+\int^{g_-\left(g_-^{-1}(x_0)-\tau_0\right)}_{x_0}\left(\ec+\pc\right)\left(g^{-1}_-(x)-g^{-1}_-(x_0)+\tau_0,x\right)\,dx\nonumber\\
&&\quad\leqslant M_0^{\mathrm{in}}+\|\phic^-\|_{L^2},\quad \forall\,(\tau_0,x_0)\in\Omega_4\label{uela5}.
\end{eqnarray}
Combining (\ref{uela1}), (\ref{uela3})--(\ref{uela5}) with (\ref{estphi1})--(\ref{estpsi2}) conculdes the proof of the proposition.\prfe

\subsection{A local existence theorem}\label{local}
The existence of solutions in the class $\mathcal{C}_T$, for some $T>0$, can be proved by standard methods. Moreover, the solution can be continued in a larger time interval as long as the momentum support of the particle density is bounded. Precisely, we have
\begin{Theorem}\label{localexistence}
Let an admissible data set $\{\fcin,\phicin,\psicin,\phic^-,\psic^+\}$ be given. There exists $T>0$ and a unique $(\fc,\Uc,\phic,\psic)\in\mathcal{C}_T$ solution of (\ref{vlasova})--(\ref{maxwell2a}) such that
the conditions (i)-(ii) in the statement of Theorem \ref{main1} are attained.  Let $T_{\mathrm{max}}$ be the maximal time of existence; then 
\[
\mathcal{P}_\cap(T_{\mathrm{max}})<\infty\Rightarrow T_{\mathrm{max}}=\infty.
\]
\end{Theorem}
Given the representation formulas for the fields established in Section \ref{representationformula}, the proof of the above theorem proceeds exactly as in the case of the Cauchy problem. The argument is based upon an iteration scheme first introduced in \cite{Ba}. We refer to \cite{GS2} for details, see also \cite{KRST}. As a matter of fact, the proof in one spatial dimension is much easier and is essentialy given in \cite{GSh2}, although the result is not explicitely stated.

\subsection{Bound on the momentum support and proof of Theorem \ref{main1}}\label{boundmomentum}
Let $(\fc,\Uc,\phic,\psic)\in\mathcal{C}_T$ be a local solution which match the initial-decay data. The uniform bounds on the fields proved in Section \ref{estfield} lead to an estimate on the function (\ref{momentumsupport}). To see this, let 
\[
K=\sup_{\tau\in [0,T)}\left(\|\Uc(\tau)\|_\infty+\|\phic(\tau)\|_\infty+\|\phic(\tau)\|_\infty\right)
\] 
and define
\[
\mathcal{Q}(\tau)=\sup_{s\in [0,\tau)}\left\{\sqrt{1+|p|^2}+K\sqrt{1+x^2},\,(x,p)\in\mathrm{supp}\,\fc(s)\right\}>\mathcal{P}_\cap(\tau).
\]
Along solutions of (\ref{char1da})-(\ref{char1db}) we have
\begin{eqnarray*}
\frac{d}{ds}\left(\sqrt{1+|p|^2}+K\sqrt{1+x^2}\right)&=&\Uc\frac{p_1}{p_0}+\left(\phic+\psic\right)\frac{p_2}{p_0}+K\frac{p_1\widehat{x}}{p_0}\\
&\leqslant&K\left(\frac{\sqrt{1+|p|^2}}{p_0}+\frac{p_1\widehat{x}}{p_0}\right)=K.
\end{eqnarray*}
Integrating the previous inequality in the interval $[0,\tau]$, $\tau<T$, we obtain the estimate $Q(\tau)\leqslant C(1+\tau)$, for all $\tau\in [0,T)$. If the maximal time of existence $T_{\mathrm{max}}$ were finite, the latter estimate would imply $\mathcal{P}_\cap(T_{\mathrm{max}})\leqslant C(1+T_{\mathrm{max}})<\infty$, which is a contradiction to Theorem \ref{localexistence}. Thus $T_{\mathrm{max}}=\infty$ and this concludes the proof of Theorem \ref{main1}.

\bigskip
\noindent {\bf Acknowledgments:} The author acknowledges support
by the FCT, Portugal  
%{\it Funda\c{c}\~ao para a Ci\^encia e a Tecnologia} 
(contract SFRH/BDP/21001/2004).

\end{document}